# Simulation of Pedestrian Movements Using Fine Grid Cellular Automata Model

**{sarmady@cs.usm.my, fazilah@cs.usm.my, azht@cs.usm.my}**
Siamak Sarmady[a], Fazilah Haron[a,b] and Abdullah Zawawi Talib[a]

[a] School of Computer Sciences, Universiti Sains Malaysia, 11800 USM, Penang, Malaysia.
[b] College of Computer Science and Engineering, Taibah University, Madinah, Saudi Arabia

***Abstract –*** *Crowd simulation is used for evacuation and crowd safety inspections, study of performance in crowd systems and animations. Cellular automata has been extensively used in modelling the crowd. In regular cellular automata models, each pedestrian occupies a single cell with the size of a pedestrian body. Since the space is divided into relatively large cells, the movements of pedestrians look like the movements of pieces on a chess board. Furthermore, all pedestrians have the same body size and speed. In this paper, a method called fine grid cellular automata is proposed in which smaller cells are used and pedestrian body may occupy several cells. The model allows the use of different body sizes, shapes and speeds for pedestrian.*

*The proposed model is used for simulating movements of pedestrians toward a target. A typical walkway scenario is used to test and evaluate the model. The movements of pedestrians are smoother because of the finer grain discretization of movements and the simulation results match empirical speed-density graphs with good accuracy.*

## 1 Introduction

Crowd simulation models can be used to predict crowd efficiency and performance issues in the design of buildings and public facilities. It is also possible to estimate evacuation time during emergencies using crowd simulation models. Simulation of crowd and pedestrian movements may also be used in producing animations.

Continuous methods for crowd simulation like force based models produce smooth movements and could potentially produce accurate and realistic results. However these models use complex mathematical rules. Adding features to such models requires modifications of the relatively complicated differential equations. In addition, solving such systems of equations using numerical methods is computationally expensive and modification of models may require modification of numerical methods as well.

Cellular automata models on the other hand use simple and flexible transition rules to identify the cell to which a pedestrian will move in the next step. Regular cellular automata models for crowd applications [1-5] divide the movement space into relatively big grid cells. Each cell accommodates a pedestrian. At each simulation time step, pedestrians move into one of the neighbouring cells or stay where they are. The big size of the cells and the discrete nature of the model dictate that the same size and equal speeds are used for all the pedestrians. However, in real world scenarios, pedestrians have different sizes and their movement speeds could be very different. In this paper we intend to improve the accuracy of cellular automata models while keeping their advantages. We use a method called fine grid cellular automata to simulate least effort straight line movement toward a target.

Different movement patterns can be observed in pedestrian movements. However majority of the pedestrian movements can be described in terms of movements toward successive targets. The concept of least effort [6] path selection can be observed in pedestrian movement behaviour in the mentioned situations. In a walkway for example, pedestrians walk in an almost straight path towards the end of the walkway and sometimes exit somewhere in the middle. The concept of least effort mostly results into a shortest path straight line walking toward the target. In addition to least effort movement behaviour, pedestrians show other behaviours like collision avoidance, density and congestion aversion and group formation. In this paper, we first present basics of our fine grid cellular automata model and then we formulate a variation of least effort movement model for this kind of cellular automata. In a previous work [7] we introduced a multi-layer model of human actions and movements. We used discrete event simulation methods to simulate actions of the pedestrians. A macroscopic movement layer would simulate the navigational large scale movement behaviours and a cellular automata movement layer was used to simulate microscopic small scale movements of pedestrians. The new model could replace the older cellular automata model in order to allow more realistic simulation of pedestrian movements.

The organization of the paper is as follows: Section 2 reviews the existing related work. In Section 3 we introduce the fine grid cellular automata model. In Section 4 we present the simulation results and evaluation of the model. We conclude the paper in Section 5.

## 2 Related Work

Collective movements of individual pedestrians form the behaviours and movements of crowd. Therefore, one approach toward modelling crowd could start from creating a model that simulates movements and behaviours of individual pedestrians. Large scale behaviours of pedestrians like way finding and navigational decisions can be dealt in a separate model from the small scale behaviours such as collision avoidance during a walk. However, some of the existing models combine both behaviours into a single model [2]. Navigational and way finding behaviours are referred to as macroscopic movements while small scale behaviours are called microscopic movements. In contrast to “microscopic and macroscopic behaviours”, a separate categorization divides crowd models into “macroscopic, mesoscopic and microscopic models”.

In macroscopic models like Hughes’ continuum theory of pedestrian flow [8], the crowd is considered as a single liquid like entity with varying densities and speeds in its different parts. In these models the overall properties of the crowd, such as flow, density and speed in different parts of the system are estimated using differential equations [8-13]. Such calculations are easier and less compute intensive because they do not involve simulating thousands of pedestrians. Mesoscopic models consider the individual pedestrians but their movements are not simulated. Lovas [14] modelled the crowd using discrete-event and queuing network methods. In this model, the crowd system is represented by a network of nodes (rooms) and links (doorways). The capacity of doorways and corridors is used to calculate flow and to predict congestion and queuing. The model considers the way selection of individual pedestrians but the information is only used to estimate flow in each path and, as mentioned, the movements are not simulated. Another example of mesoscopic models is the work of Hanicsh et al. [15]. These models provide the number of pedestrians and average density, flow and speed in a region but details about the different points inside a region are not provided. If detailed information like density and speed for different points in an area are needed these models could not be used. Microscopic models on the other hand model behaviours and movements of individual pedestrians in crowd and the emerging behaviour of those pedestrians will simulate the behaviours of the whole crowd. Microscopic models allow users to observe more details about behaviours of pedestrians in different situations (e.g. emergency evacuation), their interactions with obstacles and building elements; as well as flow, density and speed. Higher computing power requirement is of less concern because more powerful multi-core computers are available nowadays.

A few dominant approaches have been used to model microscopic movements. Physics based methods use laws of physics to model movements of pedestrians. Attraction and repulsion forces are used in Teknomo’s forces model [16], Helbing’s social forces [17] and Okazaki’s magnetic forces method [18]. However, each of the three models uses different physical concepts and rules to model interactions of pedestrians. Helbing et al. [17] proposed a model which describes the movements in a crowd in terms of imaginary forces being applied to pedestrians. The movements are then calculated as if pedestrians are particles with specific mass subjected to physical attraction and repelling forces. These forces include the repelling force between pedestrians which represents the tendency to avoid collisions, the repelling force between pedestrians and obstacles (e.g. walls) which represents the pedestrians avoiding collision with walls and obstacles, and the attraction force between pedestrians and target points which represents the inclination of pedestrians walking towards their target. Since these forces are not physical rather they symbolize the internal motivations, they have been called as “social forces”.

Cell-based methods divide the movement space into grid of cells and use cellular automata [1-5], distance maps [19, 20] or other methods to calculate transitions of pedestrians between the cells. Cellular automata models belong to a category of models which divide the simulation area into cells of a grid. In terms of cell occupation, cellular automata models can be categorised into three groups. In the first and most common category, no more than one pedestrian can be present in each cell. Since applying this rule implies that pedestrians cannot overlap each other, collision is automatically avoided. Furthermore in conventional cellular automata models, each pedestrian spreads and occupies only one cell. The second category of models allows more than one pedestrian in each cell [21, 22]. These models can be considered as a hybrid between macroscopic and microscopic (i.e. mesoscopic) models. Each big size cell in this category can contain tens of pedestrians. These kinds of models lack enough details about the movements of individual pedestrian. In the third category, each pedestrian occupies more than one cell. This research proposes a model which fits the latter category. In the model, each pedestrian may occupy and spread to several cells. The collision is still avoided between bodies of different pedestrians.

Pedestrian transition to neighbouring cells is based on simple rules. Cellular automata transition rule could be simple mathematical equations which determine the next transition cell for each pedestrian. The next cell is normally one of the adjacent cells. Since the rules are simple and flexible, and the calculations are performed for individual pedestrians almost independently from others (i.e. except checking the occupancy of grid cells by others), it is easy to integrate higher level behavioural models (e.g. way-finding, decisions and actions) and the simulations are fast. Even though the movement rules of individual pedestrians are simple, the collective and emergent behaviours of the crowd of pedestrians could be complex and sometimes not predictable. This complexity is one of the reasons the simulations are performed in the first place. Several crowd simulation models have used the cellular automata approach. Nagel-Schreckenberg [23], Biham [24] and Nagatani [25] used cellular automata to model movements of vehicles in streets and highways and then they were applied to pedestrian movements. Among the other researchers who have used cellular automata for the simulation of pedestrian movements Dijkstra [3], Blue and Adler [1] and Kirchner, Schadschneider and Burstedde [2, 4, 5] can be mentioned.

Another major crowd simulation approach uses high level behaviours that can be observed in the crowd. Rule based model uses behavioural rules to simulate crowd of simple creatures like flocks of birds, group of fishes and herds of animals [26]. The model has also been used to simulate movements of groups of pedestrians. However, the microscopic behaviours of the movements (e.g. collision avoidance, density effects, behaviours at narrow passages) are more suitable for making animations than scientific simulations. Creatures in the model behave based on their perception of the other group members and the surrounding environment. Three behavioural rules are used in the model. Creatures in the flock try to avoid collision with the others (collision avoidance), try to follow and match their speed with the group (velocity matching) and finally try to stay near other members of the group (flock centring).

One more category is the velocity based collision avoidance method or velocity obstacles (VO) [27-31]. In these models collision is avoided by calculating feasible velocities (i.e. speed and direction) for the pedestrians that will not cause collision in a specific time frame and if there is no collision potential they can walk with their free flow speed. The velocity obstacles model is similar to the rule based models (in which collision avoidance is not guaranteed). The models in this category use continuous movement (as opposed to discrete moves in cellular automata). In dense crowds of higher than 5-6 pedestrians/$m^2$ people tend to walk very near to each other and almost collide. Curtis et al. [32] have used this type of model to simulate dense crowds but since collisions are totally avoided by velocity adjustment, simulations might not simulate dense crowd specific phenomena like pushing and trampling. However, it might be possible to extend the model to support dense crowd features (similar to the work done by Henein [33] for cellular automata).

## 3 Models

As mentioned earlier, the main objective of this paper is to enhance and improve cellular automata based models for pedestrian movements. The focus of the improvements is on two aspects namely smoother and more accurate movements, and the ability to match different types of crowds. We propose a cellular automata model which uses smaller cells and therefore is able to simulate smaller and finer movements and allows pedestrians with different body sizes and shapes, and movement speed. We also propose a method which allows using any empirical speed-density graph for the simulations and therefore matching different types of crowd.

### 3.1 Fine Grid Model

In the majority of the popular cellular automata models like those described in [1-5], pedestrians occupy a single cell (Figure 1). In each simulation step, some of the pedestrians transit into one of their empty neighbouring cells while some stay where they were if their desired neighbouring cells are already occupied. The rules being used for the transitions are simple and have lower computational cost than solving partial differential equations in social forces model. However, because of the coarse grain discretization of the space, movements of pedestrians are not smooth and realistic. The movements look like the movements of pieces on chessboard. They jump a large distance in each step (i.e. 0.4-0.56m in a 0.4x0.4m grid).

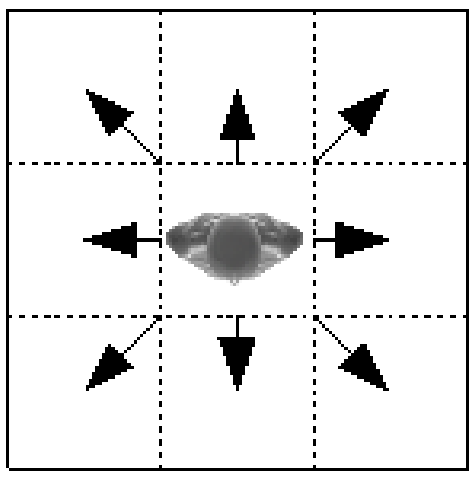

Figure 1: Transition of pedestrian into neighbouring cells in regular cellular automata

In addition to unsmooth movements, pedestrians can have just a few speed levels. If each cell is of 40cm x 40cm in size, pedestrians are able to move 0, 1, 2, 3 or 4 cells in a time unit which are equal to speeds of 0, 0.4, 0.8, 1.2 and 1.6 m/s respectively. The speed range of 0-1.6 m/s is acceptable considering that the average speed of a pedestrian is around 1.3 m/s [34]. However, the discrete choices (i.e. 5 possible values), limit the accuracy of the simulation and make it difficult to do fine adjustments on the model with regard to pedestrian speed. This will in turn limit the capability of adjusting other model parameters such as density which are somehow related to speed. Using smaller cells (i.e. finer discretized speed steps) along with the speed control method which will be described shortly, minimizes the mentioned problem. Smaller cells allow smaller displacements in each simulation step and therefore more speed levels. Depending on the requirements, grid cells could be 5cm x 5cm or even smaller. In a 5cm x 5cm grid and 0.025 second period for each time step, 41 possible speed values (i.e. moving 0,1,…40 cells per second) can be produced.

Top view of pedestrian body is mapped to the smaller cells and therefore each pedestrian will occupy more than one cell. In the new method which is called fine grid cellular automata, pedestrians can have different body shapes, sizes and different speeds. It is also possible to create and use different profiles with different settings which represent different groups in a society (e.g. kids, teenagers and adults). It is possible to create large numbers of maps for pedestrians with different orientation and movement direction (e.g. 36 maps for 0º, 10º, 20º and 30 º…directions). However, in this research only eight maps are used for eight major directions. Creating additional maps would make the implementation more difficult without adding much benefit. For smaller sized grid cells creating more maps could result in more natural and smoother visual results. Figure 2 shows maps for four of the eight orientations. During the simulation, orientation is identified using the direction of the target point. One of the cells in the weight centre of each map is considered as the centre point. Cellular automata transition rules and next cell calculations are applied to the centre point. Other cells of the map are moved along with the centre point. Collision avoidance is done for all of the map cells. None of the map cells can overlap the map cells of other pedestrian.

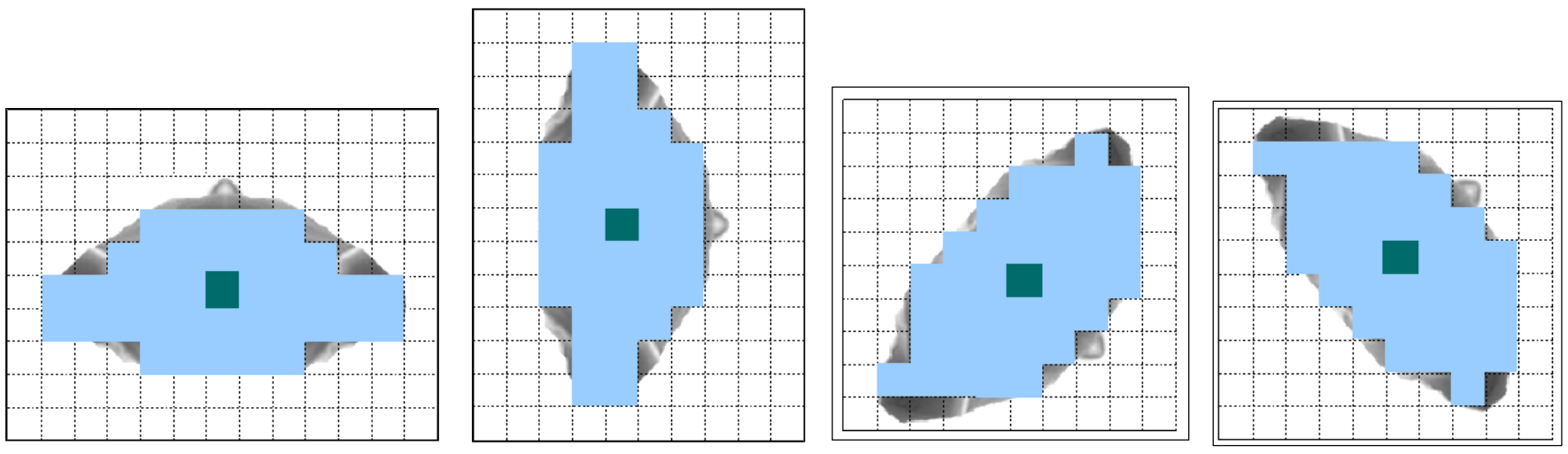

Figure 2: Sample pedestrian maps for different orientations

Figure 3 compares the maximum number of pedestrians in a square meter in both fine grid and traditional cellular automata models. As it can be seen fine grid cellular automata can accommodate pedestrian densities of up to 7-8 pedestrians/$m^2$ with the presented body map. More pedestrians could possibly be fit to one square meter if smaller maps (i.e. for kids and people with smaller body size) are used.

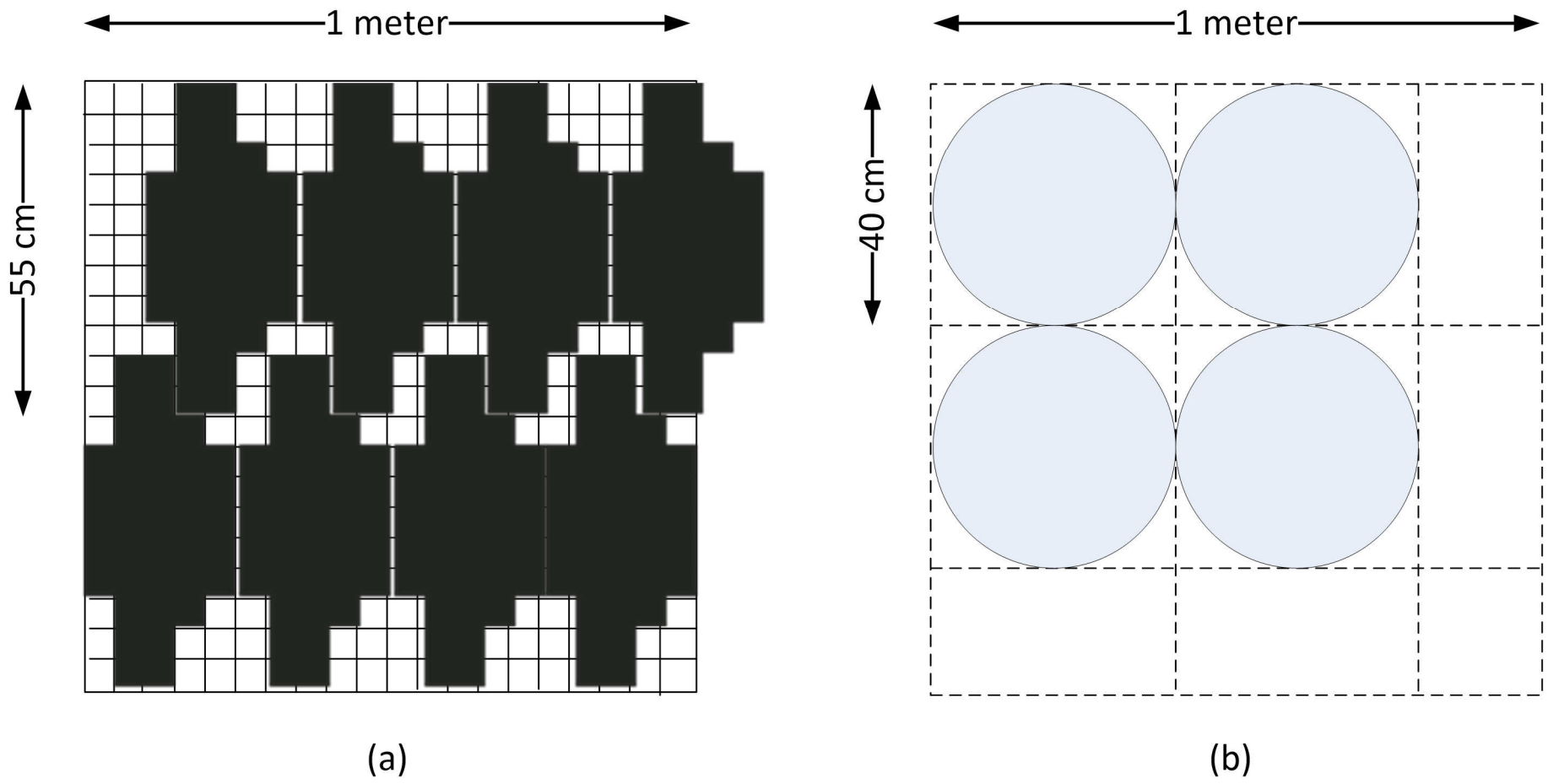

Figure 3: Maximum possible density in (a) fine grid and (b) traditional cellular automata

The algorithm used for the next cell selection (i.e. smalls scale movements) is a modified version of the least effort straight line movement algorithm [36] adapted to fine grid method. In the least effort algorithm, movement of pedestrians are calculated based on the assumptions that the desired movement path is the one which needs the lowest effort. The least effort usually means the shortest path (i.e. the direct line) between current position and the target. In each simulation step, the centre point of each pedestrian moves to one of its neighbouring cells or stays at the same cell. All the cells of the pedestrian map are moved along with the centre cell in the same simulation step. The neighbouring cell which is nearer to the target location gets selected most of the time. The movements continue until the pedestrian reaches its target cell. At this point a small-scale movement is completed and the pedestrian will proceed to the next small-scale movement toward another target. A “large-scale” or macroscopic navigational movement consists of a series of “small-scale” or microscopic movements which eventually bring the pedestrian to its final target.

## 3.2 Probability Model for Cellular Automata Transitions

Transition probability $P_i$ is the probability of the centre point of the pedestrian map moving into the neighbour i and it is given by:

$$P_i = N\,M_i \qquad (1)$$

For each neighbour i, the parameter $M_i$ is calculated using the following equation:

$$M_i = n_i\, e^{\beta \frac{R_{min}}{R_i}}$$
$$R_{min} = Min(R_i),\ n_i \in \{0,1\},\ \beta \gg 0,\ R_i \neq 0 \qquad (2)$$

N is a normalizing parameter which adjusts the sum of the probabilities for all neighbours to one. The value of N can be calculated by:

$$N = \frac{1}{\sum M_i} \qquad (3)$$

In Equation 2, $R_i$ is the distance between the cell i and the target; and $R_{min}$ is the minimum distance to the target among all neighbouring cells. $R_{min}/R_i$ therefore shows the ratio of the distance of cell i to target in relation to the minimum distance to the target among the cells. As mentioned in the conditions, the equation is only valid when $R_i \neq 0$. $R_i=0$ means the neighbour i is the target. If this is the case, the cell i gets a probability of 1 (multiplied by $n_i$ for collision detection) and all other cells get a probability of 0 and pedestrian will try to move into the cell which is on the target. Including the condition in the equation would give the alternate form:

$$R_{min} = Min(R_i)\,,$$
$$M_i = \begin{cases} n_i\, e^{\beta \frac{R_{min}}{R_i}} & R_{min} > 0 \\ n_i & R_{min} = 0, R_i = R_{min} \\ 0 & R_{min} = 0, R_i \neq R_{min} \end{cases} \qquad (4)$$

The basic concept behind the calculation of $M_i$ values is that the probability of selection for each neighbour is inversely proportional to its distance to the target. If the distance is shorter, the cell gets a higher probability and vice versa. However, the difference between the distances to the target could be very small for the neighbouring cells. The small difference between the distances will cause the probabilities to be very near to each other. The cell nearest to the target should be selected most of the time and the cell farther from the target should be selected very rarely. In order to increase the difference between probabilities for the eight neighbouring cells an additional parameter $\beta$ is used as exponent.

If moving into the i-th neighbour of the centre point would cause collision, entering into that cell should be avoided. This is taken care of by the parameter $n_i$. All the j-th cells of a pedestrian map will be checked to make sure that none of them collides with an obstacle or another pedestrian's map. Collision of j-th cell of the map (i.e. $l_j$=1) would cause the (1-$l_j$) become zero. Therefore, $n_i$ will become zero for the i-th neighbouring cell and the probability of moving into that neighbour will be zero.

$$n_i = \prod_j (1 - l_j) \qquad (5)$$

Using very large values for the parameter $\beta$ (i.e. $\beta \to \infty$), will cause the nearest neighbouring cell to the target to be selected all the time. The extreme case could be shown using the following equation.

$$P_i = \begin{cases} 1 & R_i = R_{min} \\ 0 & else \end{cases} \qquad (6)$$

### 3.3 Simplified and Faster Probability Model

Probability model presented in previous section is computationally expensive because it involves floating number exponentiation and several multiplications. There is, however, one more important drawback. In long distances to the target, the difference between the values of $R_i$ and $R_{min}$ (the smallest $R_i$ among the neighbours) are very small. In order to give a higher probability to the nearest cells to the target, bigger values for $\beta$ should be used in Equation 2. That in turn will require the use of programming variables with larger capacity and slower calculation speed. Considering that millions of these calculations may take place in every simulation step, a better and less expensive method would be beneficial.

$$\forall R_i : R_i \gg 0 \;\Rightarrow\; \frac{R_{min}}{R_i} \simeq 1 \qquad (7)$$

Based on the fact that cells which are nearer to the target get selected most of the time, we may use a random number with a suitable distribution to do the same. The probabilities of moving into neighbouring cells will not be accurately in proportion to their distance with the target but the distribution parameters can be adjusted to sufficiently reproduce the empirical data. $M_i$ for each of the neighbouring cells is calculated by simplifying Equation 2 to the following form:

$$M_i = n_i \frac{R_{min}}{R_i} \tag{8}$$

$$R_{min} = Min(R_i),\ n_i \in \{0,1\}, \quad R_i \neq 0$$

In the next step, the neighbours are sorted based on their $M_i$ value in descending order. Since the $M_i$ is directly proportional to the desirability of a cell, the cell with higher $M_i$ and better rank (i.e. smaller index) should be selected most of the time. Poisson statistical distribution (with a $\lambda \leq 1$) can be used to randomly select the next cell.

Since the new approach uses the ranking of the neighbours based on their distance to the target, Equation 8 can be further simplified into the following form, which reduces large number of calculations.

$$M_i = \frac{n_i}{R_i} \tag{9}$$

$$n_i \in \{0,1\}, \quad R_i \neq 0$$

The proposed model uses a randomized order of movement for pedestrians and performs the movements in a single cycle. This will guarantee that pedestrians will not perform their moves always in the same order (which would be unrealistic). Furthermore an additional mechanism of density perception is used to control pedestrian speed and movement direction.

### 3.4 Speed-density Model

In most conventional cellular automata crowd models, simulated pedestrians will move with their full free flow speed unless they arrive at an obstacle or another pedestrian. This unrealistic behaviour will affect the visual outcome as well as the quantitative results of the simulation. Therefore a model which uses perception of density to adjust the speed of pedestrians has been used.

Based on the fact that density affects the speed of pedestrians and that pedestrians adjust their speed based on perception, the proposed model uses density perception as the base for speed adjustment. The local density is measured for a rectangular area along the pedestrian's walking path. The forward local density is then used in the speed adjustment algorithm. The local perception area size is an adjustable model parameter. The size represents the area which affects pedestrian speed adjustment decisions and therefore its length is likely to be in the range of a few meters.

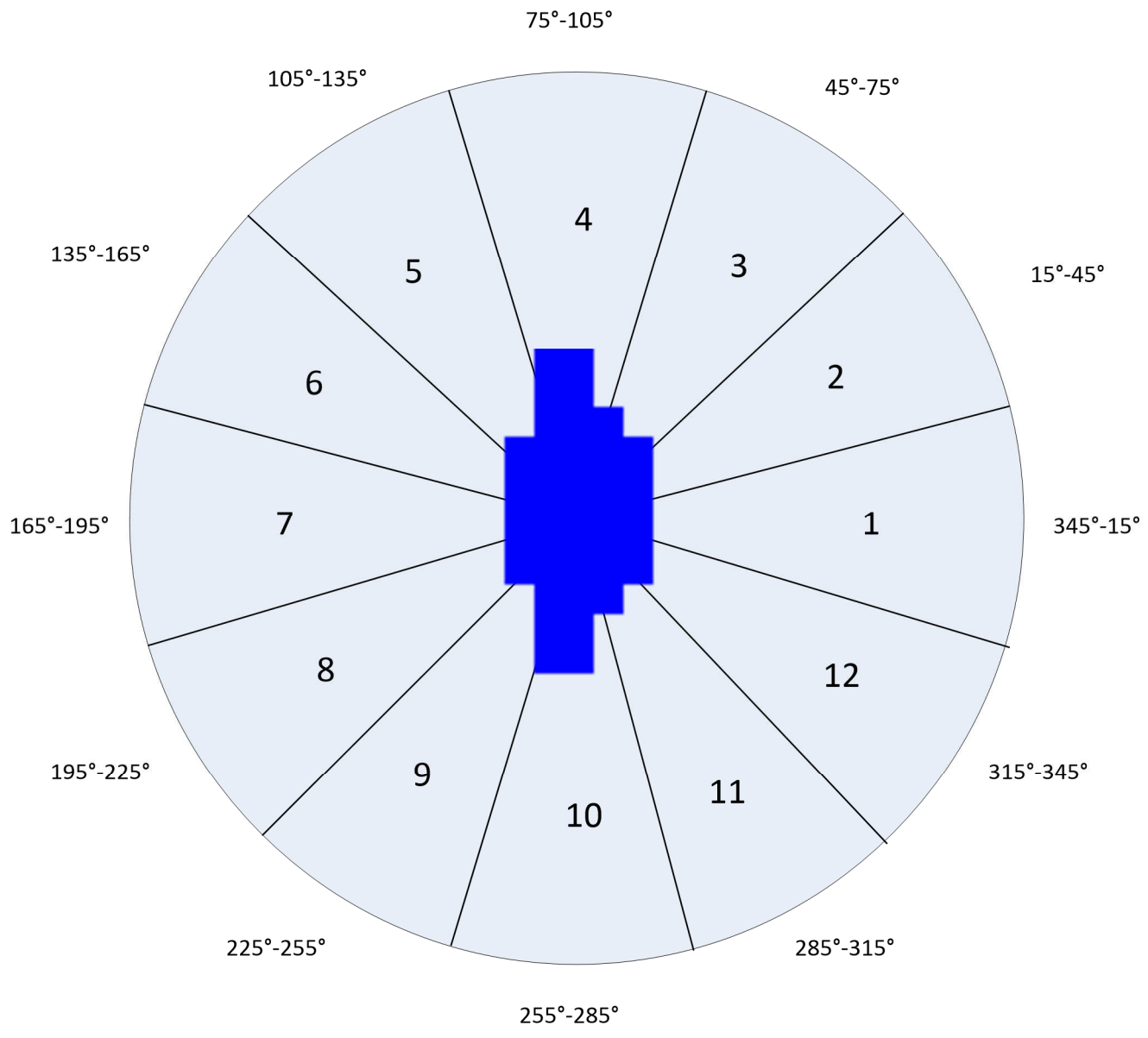

Figure 4: Pedestrian movement heading

In our implementation we divide the possible heading directions into twelve sectors and determine the density calculation (i.e. perception) area based on the sector of the heading (Figure 4). A rectangular area along the movement direction is used for density measurement. Figure 5 shows the density rectangles for sectors 1, 4, 7 and 10 in Figure 4.

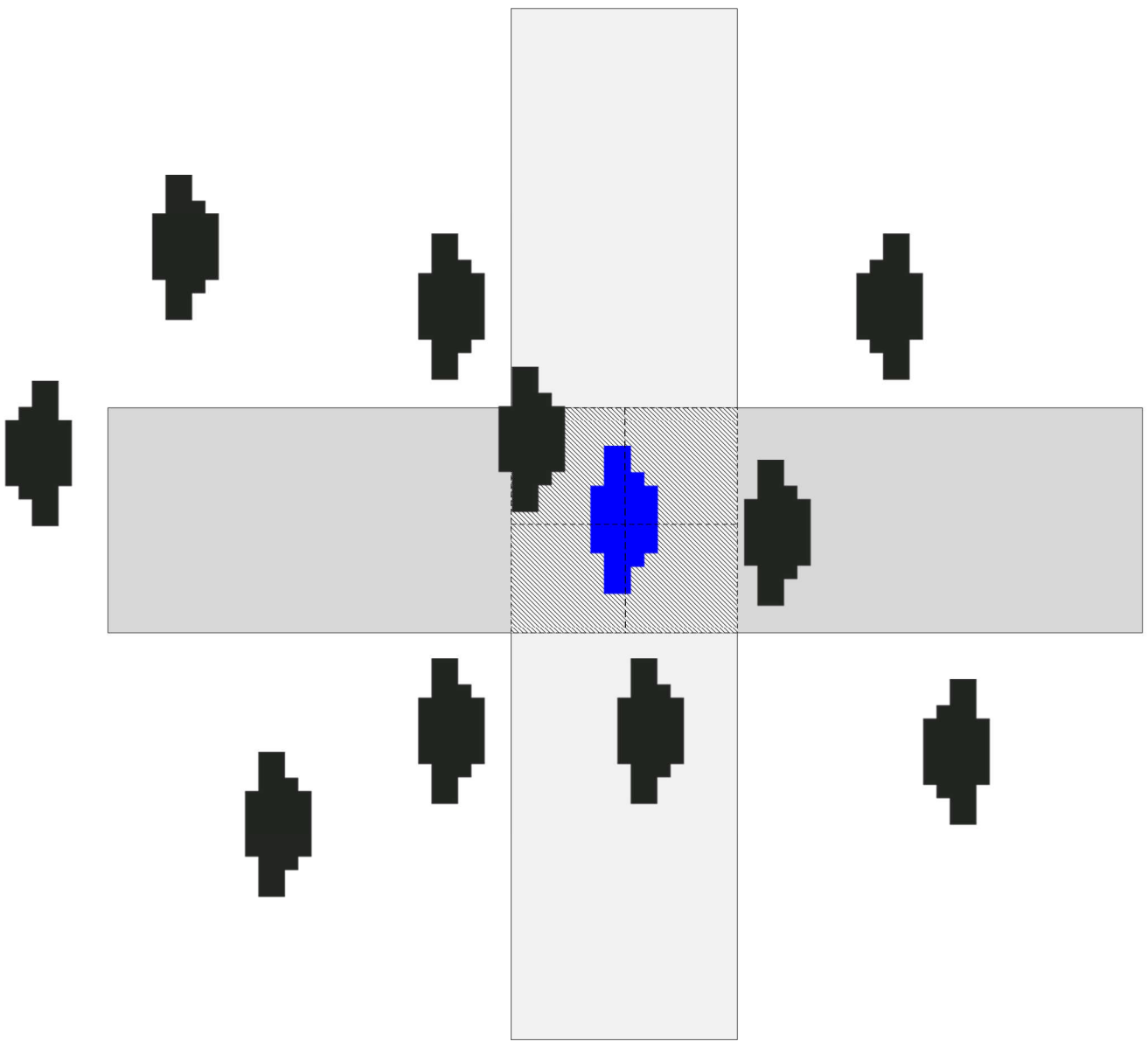

Figure 5: Perception rectangles for different headings: Sectors 1, 4, 7, 10 of Figure 4

In regular cellular automata crowd models the relation between density and speed is an outcome of the equations, parameters and settings of the model. Researchers adjust and calibrate model settings in such a way that the speed-density graph roughly matches existing empirical graphs. This sometimes proves to be very tricky and difficult because of numerous model parameters. That makes it difficult to find settings which provide the best results. In addition, non-fine grain size of the cells in regular cellular automata models limits the accuracy of speed adjustments.

A new active and adaptive approach is used to tackle the problem of calibrating simulations with empirical graphs. Perception information (i.e. local density around the pedestrian) is used to adjust the speed of each pedestrian in such a way that it matches the empirical graph at any moment. Based on the local density, a matching speed value is extracted from the desired empirical speed-density graph.

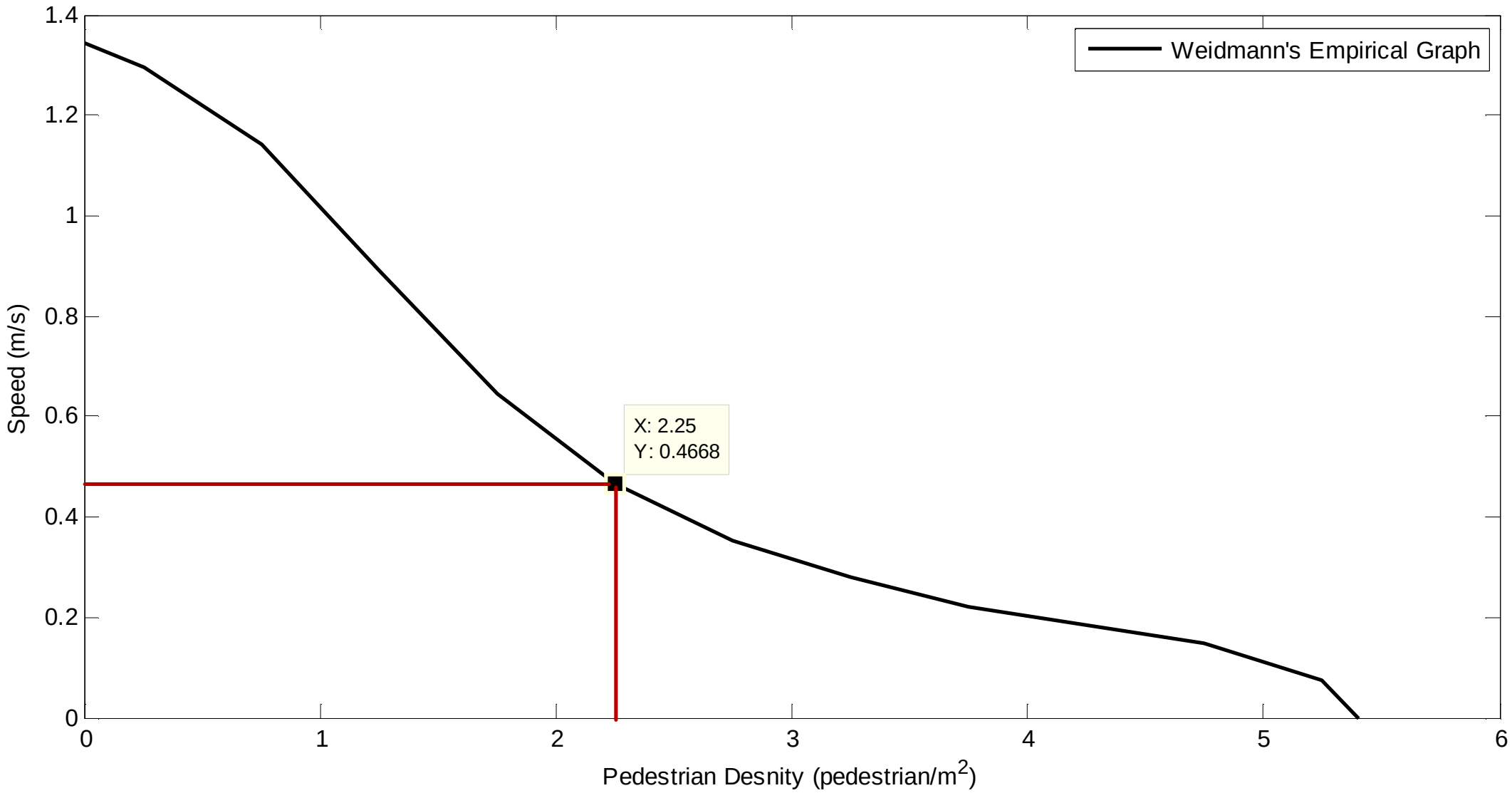

Figure 6: Matching speed for a specific density value

Figure 6 shows how this is done using Weidman's empirical data for a pedestrian who perceives a density of 2.25 pedestrians/m$^2$. The graph specifies that the pedestrian speed should be around 0.46 m/s for this specific density. The empirical speed-density graphs are available in the form of a two dimensional array of speed-density data. If a specific density value is not in the array, the speed value is found using linear interpolation. The equation below shows how the interpolation is done. In the following equation $\rho_1$ and $\rho_2$ are the two existing values in the array and $\rho$ is the desired density where $\rho_1 < \rho < \rho_2$ and $V1 < V < V2$. V is the speed of the pedestrian in a crowd with a density of $\rho$.

$$V(\rho) = V_1 + (V_2 - V_1)\,\frac{\rho - \rho_1}{\rho_2 - \rho_1} \tag{10}$$

As discussed earlier, the free flow speed varies for different pedestrians according to normal distribution. Based on the distribution, a free flow speed is randomly determined for each of the pedestrians. The speed value extracted from the speed-density graph is then adjusted to the specific pedestrian using the following equation.

$$V_p = \frac{V_{fp}}{V_{max}} \tag{11}$$

In the equation, $V_p$ is the maximum speed of the specific pedestrian, $V_{fp}$ is the free flow speed of the pedestrian which is randomly calculated from the normal distribution and $V_{max}$ is the maximum speed on the empirical graph (i.e. speed for the density=0). The calculated speed is then forced to the specific pedestrian in the simulation. It is expected that by forcing the speeds-density graph values to simulated pedestrians, almost the same speed-density graph will be obtained for the whole crowd in the simulation results.

The speed control is done by limiting the number of cells each pedestrian travels in a second. For example, if the size of the grid cells is 5cm x 5cm, the maximum moves in each second is 40 cells, the speed of pedestrian is supposed to be 1.3 m/s, and its path is parallel to grid lines (i.e. not diagonal), it would mean the pedestrian should only travel 26 cells in one second. The 26 direct moves are randomly distributed among the 40 simulation steps in each second (Figure 7). The 'X' marks show the turns in which a move is performed and the empty cells represent the turns in which no moves happen.

| | | | | | | | | | | | | | | | | | | | | | | | | | | | | | | | | | | | | | | | |
|---|---|---|---|---|---|---|---|---|---|---|---|---|---|---|---|---|---|---|---|---|---|---|---|---|---|---|---|---|---|---|---|---|---|---|---|---|---|---|---|
| X | | X | X | | X | X | X | | | X | X | | X | | X | | X | | X | X | | | X | | | | X | X | X | X | | X | X | | X | X | | X | X |

Figure 7: Random distribution of moves among 40 possible moves in a second

In each step the total performed displacement ($R_P$), remaining displacement ($R_R$), maximum remaining moves ($M_M$) and maximum moves allowed ($M_A$) are calculated. Then a binary random decision (of Bernoulli distribution) is used to determine whether in the next simulation step a move should be performed or not (as in Figure 7). By using this method the desired speed is applied to each pedestrian.

In order to use the Bernoulli distribution and generate random "move-don't move" speed control decisions, the Bernoulli probability needs to be calculated. Maximum remaining moves ($M_M$) is calculated by deducting performed moves in the current one second period (n) from the number of run steps or possible moves in a second (T) which is 40 possible moves in a second in our example.

$$M_M = T - n \tag{12}$$

Maximum allowed moves ($M_A$) is then calculated by dividing the remaining allowed displacement of the specific pedestrian in the current one second period ($R_R$) to the size of the cell (C). Notice that there is no differentiation between diagonal and direct moves in the calculation of maximum number of moves. However, in each step the total displacement is calculated and the error rising from this compromise is minimized.

$$M_A = \frac{R_R}{C} \tag{13}$$

Based on the number of allowed moves and possible moves the desired binary decision probability P is calculated as follows:

$$P = \frac{M_A}{M_M} \quad (14)$$

After calculating the Bernoulli probability a random number generator with the distribution is used to determine the “move-don’t move” decisions in each time step for each pedestrian:

$$X \sim Bin(1, P) \quad (15)$$

## 3.5 Discretization Error

Cellular automata models provide limited choice of movement in each simulation time step. Each pedestrian can move into one of its eight neighbouring cells or stay where it is. However, the displacement of the pedestrian is different when moving into the four direct neighbours than the four diagonal neighbours (Figure 8). Kretz et al. [35] have discussed the issue of discretization error in regular cellular automata and its effect on pedestrian speed. They suggested a method to select cells which give optimal speed values that produced the least error from a desired movement speed. In a previous work [7], a method was proposed to cap the movements of each pedestrian in each second of the simulation if the combination of the movements results into a displacement bigger than the free flow speed of the pedestrian. Our speed control method is basically using the speed cap method.

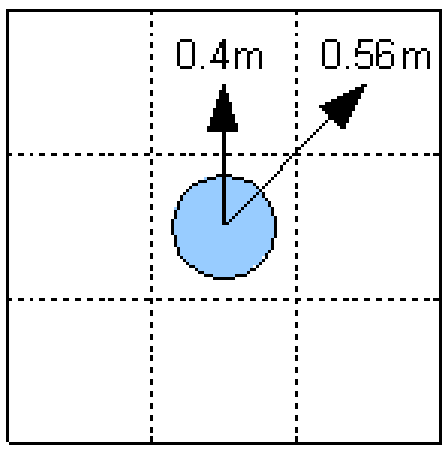

Figure 8: Displacement difference between main directions and corners

As described in previous section, the displacement of pedestrian is measured in every step and the speed is adjusted. As a result, the speed error is limited to a maximum of the diagonal size of a cell in each second. The major reason for the improvement in the accuracy of fine grid model originates from the fact that it uses smaller movement steps. Since the displacement in each time step is smaller, it is possible to adjust the displacement in a time unit (i.e. speed) with more accuracy (Figure 9). Assume that the desired speed of an agent is 1.4 m/s. Depending on the cell size and movement direction, the pedestrian could be moved over certain number of cells. The displacement should be as close to the desired displacement as possible but not exceeding it (Table 1). It can be observed that the error in a finer discretised grid is lower.

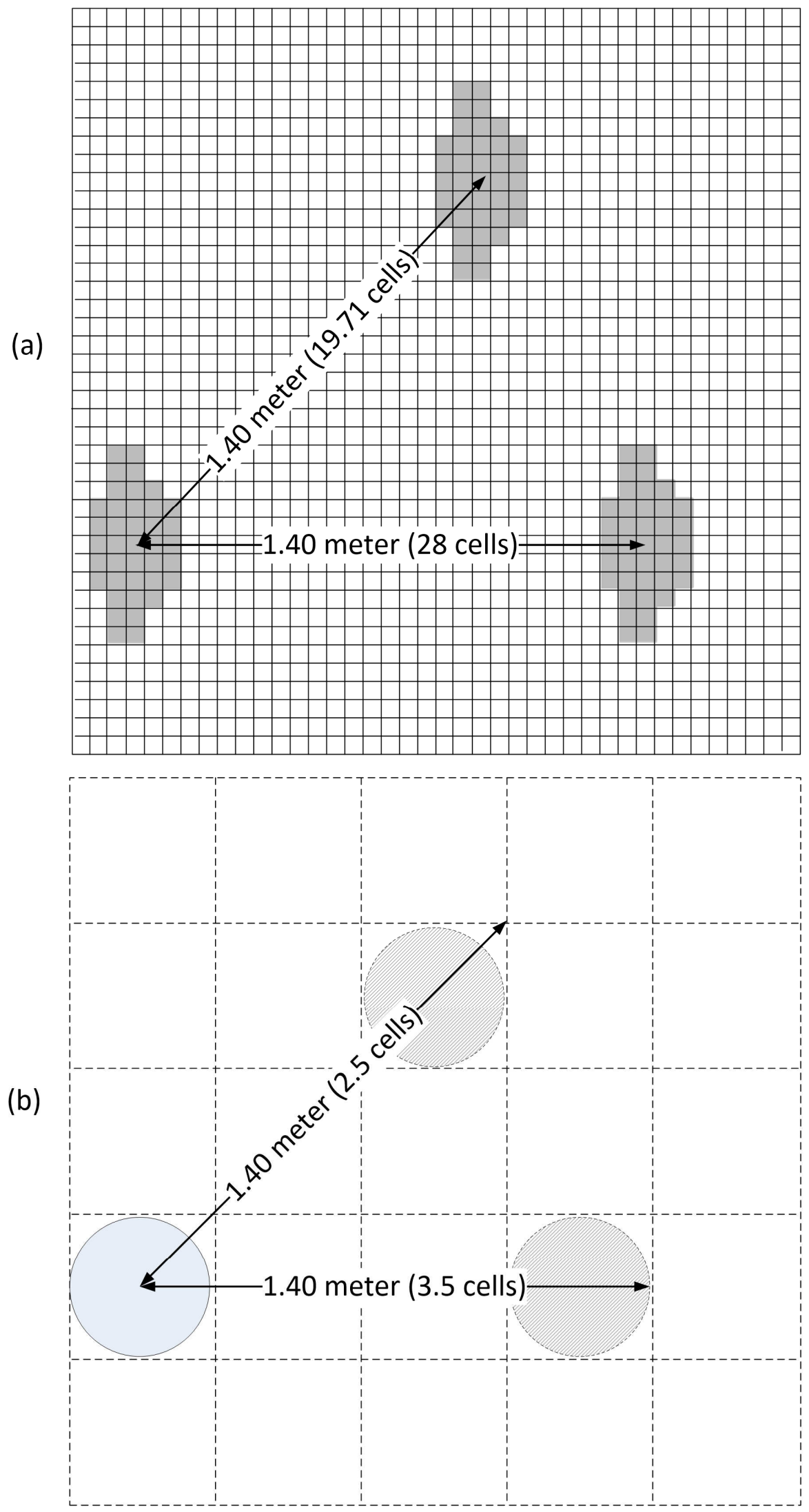

Figure 9: Discretization error due to discrete step movement in (a) 5cm fine grid cellular automata and (b) 40cm regular cellular automata

Table 1: Discretization error in regular and fine grid cellular automata

| | **Regular cellular automata (40x40cm)** | | **Fine grid cellular automata (5x5cm)** | |
|---|---|---|---|---|
| | **Direct Move** | **Diagonal Move** | **Direct Move** | **Diagonal Move** |
| **Desired Movement** | 1.4 m | 1.4 m | 1.4 m | 1.4 m |
| **Displacement of One Move** | 0.4 m | 0.56 m | 0.05 m | 0.071 m |
| **Capped Discrete Movement** | 3 cells (1.2 m) | 2 cells (1.12 m) | 28 cells (1.4 m) | 19 cells (1.34 m) |
| **Error** | 0.2 m | 0.28 m | 0 m | 0.06 m |
| **Error %** | 14 % | 20 % | 0 % | 4 % |

Furthermore, considering the size of each single move (i.e. jump) in each simulation steps on 40cm x 40cm and 5cm x 5cm grids (Figure 9), it is obvious that the movements on a finer grid are smoother and more natural. A grid with even smaller cells could produce visual results comparable to continuous models.

## 4 Simulation Results

One of the hassles of crowd simulation models including cellular automata model in our previous research [7, 36] was the number of adjustable model parameters. Excessive parameters make the model complex and difficult to adjust and calibrate. The combination of several parameters creates a huge search space from which values that produce reasonable accuracy in matching the empirical data should be found.

The speed control mechanism of the new model (i.e. fine grid cellular automata) has less adjustable parameters due to its self-adjusting nature. The smaller number of parameters (i.e. width and length of the perception area) allows investigating the smaller space of possible values easier. The self-adjusting speed control mechanism also helps to reach good accuracy in matching the empirical data.

## 4.1 Optimal Size of the Density Perception Area

As described earlier, the fine grid cellular automata model with adaptive speed control does not have large number of parameters to be adjusted and calibrated. The two parameters which need adjustment are the width and length of the perception area (i.e. the area that each agent uses to measure its perceived density). In this section, an attempt is made to calibrate the fine grid cellular automata to match Weidmann's empirical speed-density graph [37]. An optimal size of density perception area is found which minimizes the error of simulation speed-density graph compared with the desired empirical speed-density graphs. In order to find the optimum width and length of the perception area, a range of 0.5-3.5 m for both width and length is checked. The range is divided into 0.5 m increments (i.e. 0.5, 1.0, 1.5, 2.0, 2.5, 3.0 and 3.5 m). Therefore, there are 7x7 combinations of width and length to be investigated. For each perception area size, a 2000 seconds simulation is run, in which the demand starts from 1 pedestrian/s and it is gradually increased to 7 pedestrians/s. The demand schedule creates a low density crowd at the beginning of the simulation but the density reaches 4-5.5 pedestrians/$m^2$. The gradual increase of density allows speed information for different densities to be gathered. The software stores the speed-density pairs gathered for each pedestrian in a specific reporting area (i.e. a less disturbed area used for gathering speed-density data from the simulation).

In order to take randomness of the experiments into account, for each of the 49 sizes of perception area, the experiment is repeated 5 times (i.e. 245 simulation sessions). The results for each run are compared to the Weidmann empirical graph values and an MSE (i.e. mean square error) is calculated for each simulation session. An average (AMSE) is then calculated for the five MSE values obtained from the five repeated experiments. In addition a standard deviation is calculated for the five MSE values. The AMSE represents how much the simulation results for the specific perception area size adhere to the empirical graph. The standard deviation shows how much fluctuation has happened between the five experiments (i.e. the consistency of results). The intention is to find the perception area with the lowest AMSE and a reasonably low standard deviation.

Figure 10 shows the error for different perception area sizes with width of 2, 3 and 3.5 m. It can be observed that for different widths (or lengths) increasing the size of the area, initially results in reduced error but when the area size exceeds 6-7 $m^2$, the error starts to increase again. So the minimum point of error is somewhere between 6-8 $m^2$ depending on the width of area. The length and width of the area are also important. Areas of 3.5m (L) x 2.5m (W) and 2.5m (L) x 3.5m (W) are both of the size 8.75 $m^2$ but the errors they produce are different. A perception area of 3.5m (L) x 2.5m (W) produces the least amount of error and an acceptable low fluctuation of results in our simulations.

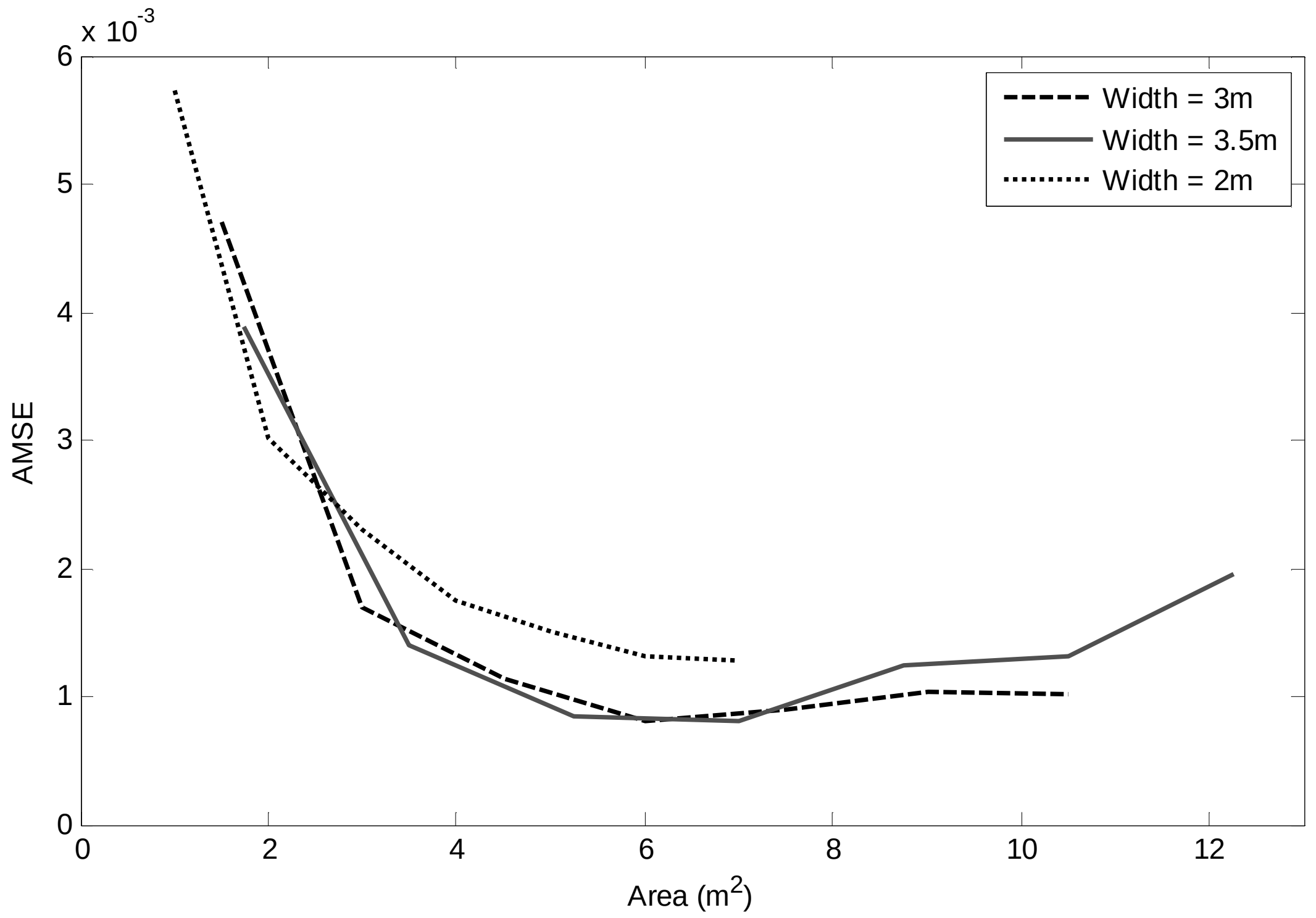

Figure 10: Error for different perception area sizes (m$^2$) and widths

Steffen et al. [38] provide an interesting discussion on the measurement of crowd parameters which includes density calculations. Measurement of density using the pedestrian/m$^2$ is prone to large fluctuations. This is because the number of pedestrians is a discrete value and increases in steps (rather than continuous increments like in liquids). In small areas, this variation might be larger most of the time. Considering slightly larger areas for the density measurement could decrease these fluctuations (Figure 11). However, considering large areas for density calculations is not always proper. If the area is homogenous and covered with almost the same density of pedestrians the results would be correct but if some parts of the area contain very dense crowd and other parts are empty, the calculated density will not represent the status of the area and any conclusions (e.g. speed-density diagrams) will be incorrect. This somehow justifies why a small area produces larger fluctuations and higher error in results.

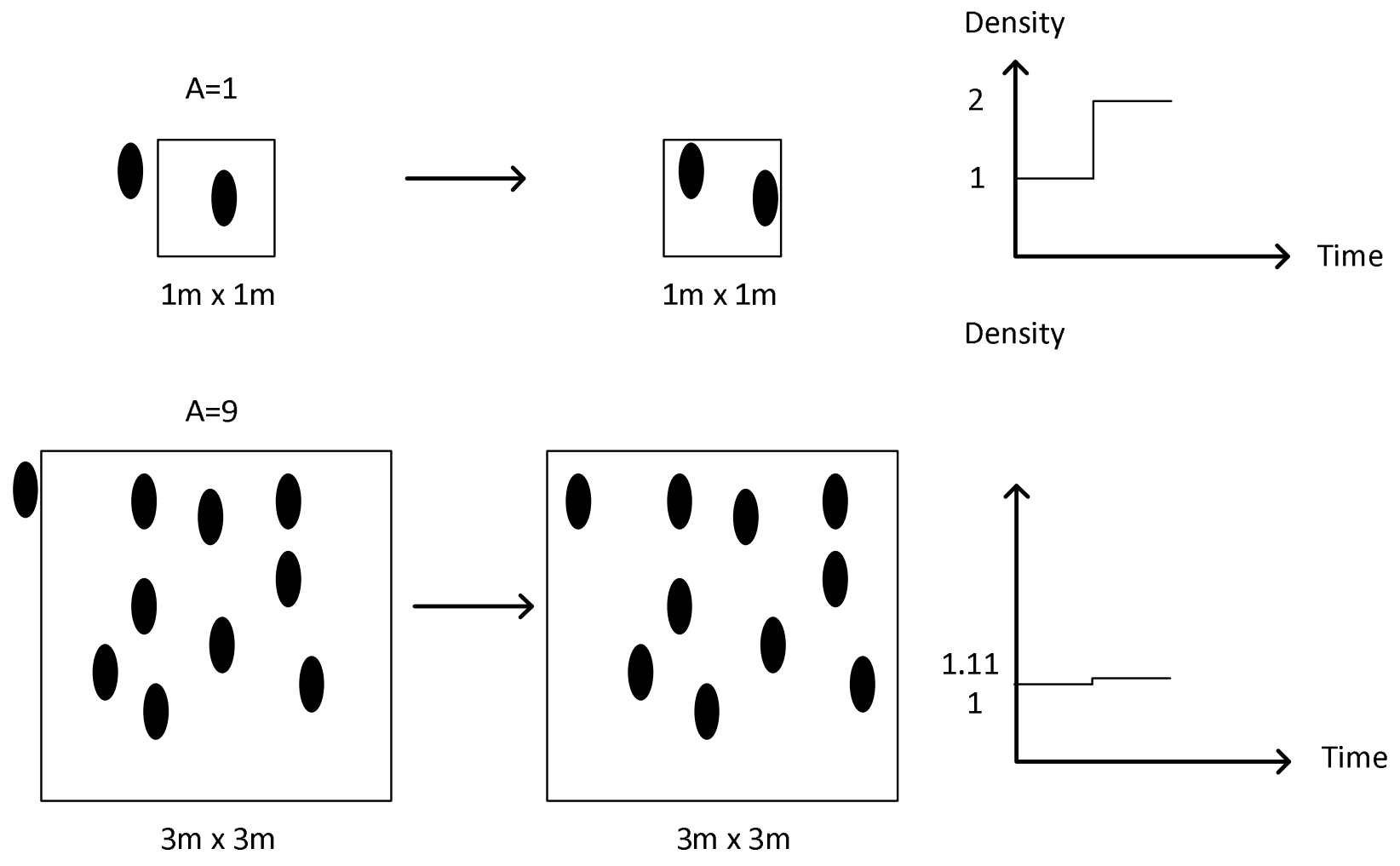

Figure 11: Fluctuation in density value due to the entry of one pedestrian into the area

## 4.2 Evaluation of the Model using Speed-Density Graph

For the experiments, we first use one of the most popular and representative empirical speed-density graphs by Weidmann [37]. However, any other empirical graph can also be used (see Section 3.4). Figure 12 shows the comparison of the fine grid cellular automata simulation results with Weidmann's speed-density data in different densities using a 3.5m (L) x 2.5m (W) perception area. It can be seen that the simulation model could match the empirical results with good accuracy. Notice that the simulation graph continues until a density value of 4.25 because that was the practical density that could be reached in the simulations of a narrow and long walkway. The specific scenario would not allow reaching higher densities. Higher density values may happen in the case of congestion or in cases where the output of the system is much lower than the input.

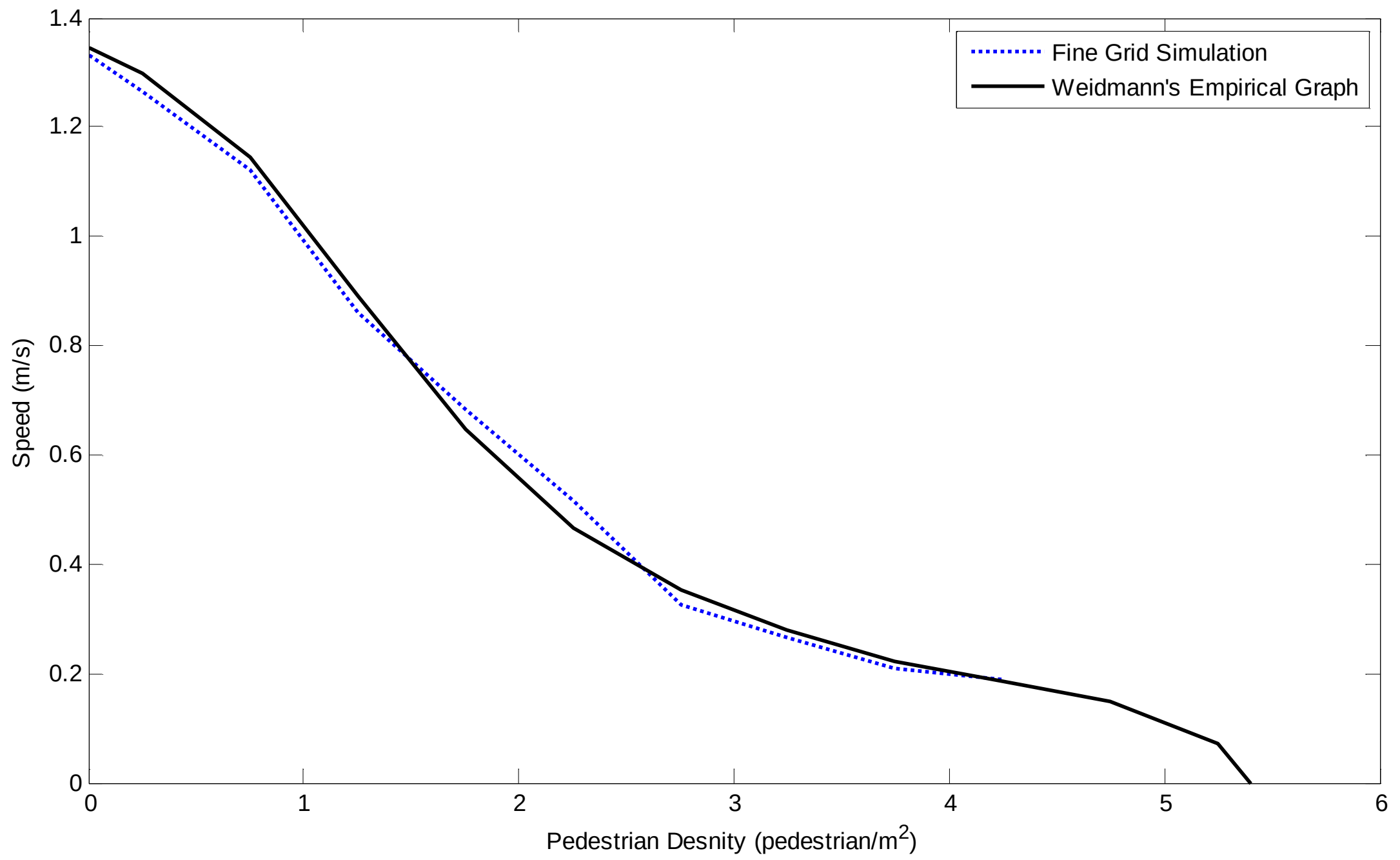

Figure 12: Comparison of the fine grid simulation results with Weidmann empirical speed-density graph

In order to show the flexibility of the model, the data by Helbing et al. [35] is also tested on the model. A curve is fitted to the Helbing's data and fed to the model. In the new experiment a 10cm x 10cm grid has been used and therefore lower accuracy is expected for matching the empirical data. After performing 245 simulations (49 perception area sizes, each experiment repeated 5 times), a perception area with a size of 2.5 m (L) x 3.0 m (W) is found to produce the least amount of error. Since pedestrian maps use fewer cells and less calculations are performed the simulation is considerably faster for a 10cm grid compared to a 5cm one. However, the experiments show that the error is at least 2 times bigger than the previous experiment with Weidmann's graph. The reason is that the 10 cm grid provides less fine grain control over the speed than the 5 cm grid and the coarser discretized speed steps result into higher accumulated error. Therefore, as one would expect, there is a trade-off between the accuracy and simulation speed. Figure 13 compares the simulation speed-density graph with the curve representing Helbing's data. The accuracy is still reasonable and it makes sense to use 10cm grid size for big scenarios in which a large simulation area and large number of pedestrians are involved.

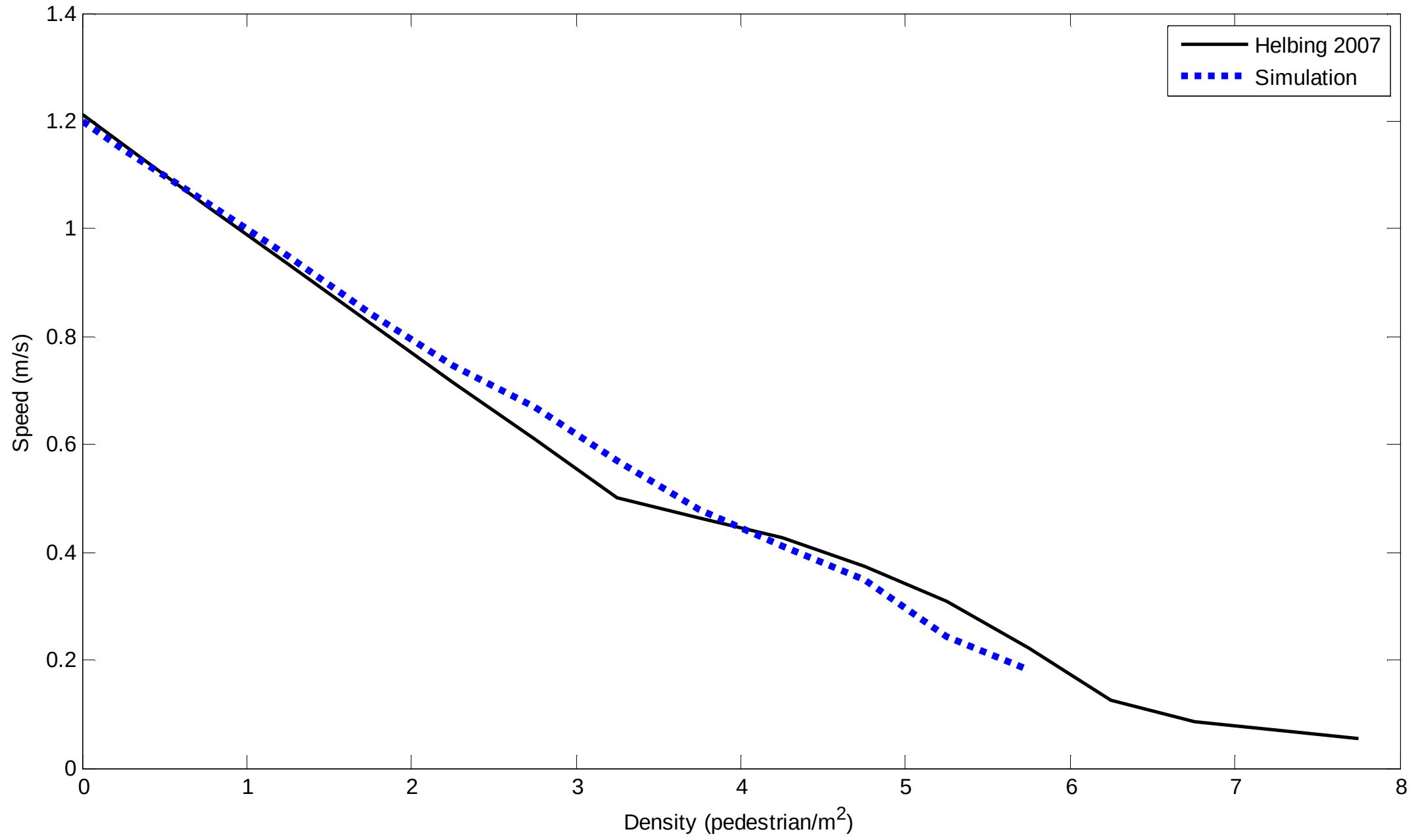

Figure 13: Fine grid simulation results vs. Helbing speed-density data

Figure 14 shows snapshots of the simulation in low to high densities. The fine grid cellular automata model is now calibrated to the Weidmann's speed-density graph with good accuracy. Considering that the model is calibrated, it can be used for the simulation of different scenarios in which pedestrians travel between a source and a destination point at different crowd densities.

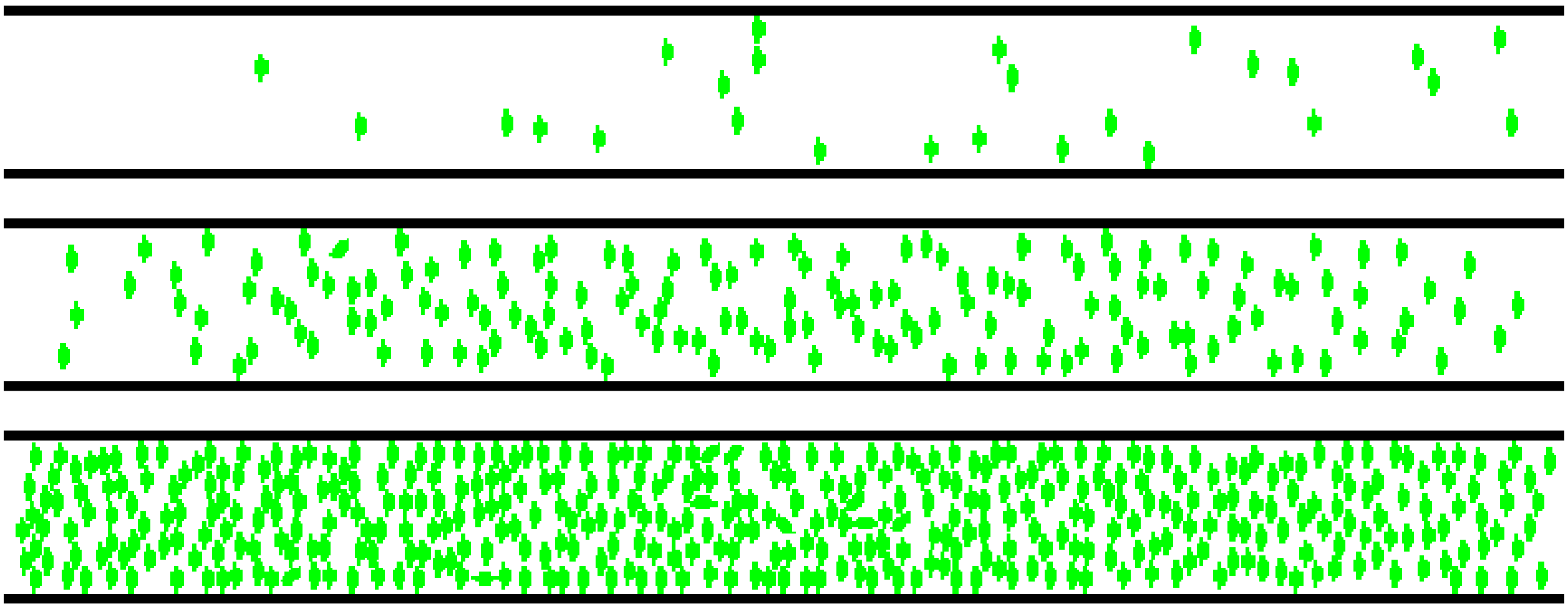

Figure 14: Simulation snapshots for low to high density crowds

## 5 Conclusion

This paper is an attempt to introduce improvements to existing cellular automata crowd simulation methods in terms of accuracy and realistic simulations. In order to have smoother movements, finer control over pedestrians' speed and their specifications such as body size and shape, a new cellular automata model called fine grid cellular automata is proposed. Smaller cells provide finer speed value selections and body maps allow different body shapes and sizes. An active speed control method allows the simulations to match the desired empirical speed-density graphs. Furthermore, since the pedestrians move smaller distances in each simulation step (e.g. 5cm or less instead of about 40 cm in regular cellular automata models), the pedestrian movements seem much smoother. The results of evaluations of the model have shown that it is able to match speed-density empirical data with good accuracy. However, the simulation speed is compromised in favour of higher accuracy in comparison to regular cellular automata models. It is possible to use slightly bigger cells to improve performance or use smaller cells to produce smoother and more accurate results.

Competition between pedestrians during emergency situations (i.e. evacuation) creates dense crowds at specific parts of a crowd system (e.g. exits). In these situations specific behaviours like pushing, falling, trampling, frictions and other behaviours can be observed specially at narrower parts like the exits. In normal mode crowd simulations, maintaining a proper speed-density relation is enough for most cases but in emergency, panicky situations the very small scale interactions between individuals are important too. Future research will attempt to adapt fine grid cellular automata to evacuation and emergency situations.

## 6 Acknowledgements

This research is supported by USM FRGS grant number 203/PKOMP/6711230 with the title “More Accurate Models for Movements of Pedestrians in Big Crowds”.